 \documentclass[12pt]{article}
 
\advance \topmargin by -\headheight
\advance \topmargin by -\headsep

\evensidemargin \oddsidemargin
\usepackage[english]{babel}
\usepackage{graphicx}
\usepackage{float}
\usepackage{comment}
\usepackage{bm}
\usepackage{amsmath,amssymb}

\newcommand{\pd}{{\partial}} 

\newcommand{\tilr}{{\tilde{r}}}
\newcommand{\tilrhO}{{\widetilde{\rho_1}}}
\newcommand{\tilrhT}{{\widetilde{\rho_2}}}
\newcommand{\txa}{{\widetilde{x_1}}}
\newcommand{\txb}{{\widetilde{x_2}}}
\newcommand{\txc}{{\widetilde{x_3}}}
\newcommand{\txd}{{\widetilde{x_4}}}
\newcommand{\txe}{{\widetilde{x_5}}}

\def\rf#1{(\ref{eq:#1})}
\def\lab#1{\label{eq:#1}}

\def\br{\begin{eqnarray}}
\def\er{\end{eqnarray}}
\def\be{\begin{equation}}
\def\ee{\end{equation}}

\def\({\left(}
\def\){\right)}

\relax

\def\tp0{\Theta_{+}^{(0)}}
\def\tm0{\Theta_{-}^{(0)}}

\def\vp{\varphi}
\def\ve{\varepsilon}

\def\u2{\mid u\mid^2}
\def\f#1#2#3 {f^{#1#2}_{#3}}

\def\win1{{\sf w_{1+\infty}}}

\def\Win1{{\sf W_{1+\infty}}}

\def\rlx{\relax\leavevmode}
\def\inbar{\vrule height1.5ex width.4pt depth0pt}
\def\IZ{\rlx\hbox{\sf Z\kern-.4em Z}}
\def\IR{\rlx\hbox{\rm I\kern-.18em R}}
\def\IC{\rlx\hbox{\,$\inbar\kern-.3em{\rm C}$}}
\def\IN{\rlx\hbox{\rm I\kern-.18em N}}
\def\IO{\rlx\hbox{\,$\inbar\kern-.3em{\rm O}$}}
\def\IP{\rlx\hbox{\rm I\kern-.18em P}}
\def\IQ{\rlx\hbox{\,$\inbar\kern-.3em{\rm Q}$}}
\def\IF{\rlx\hbox{\rm I\kern-.18em F}}
\def\IG{\rlx\hbox{\,$\inbar\kern-.3em{\rm G}$}}
\def\IH{\rlx\hbox{\rm I\kern-.18em H}}
\def\II{\rlx\hbox{\rm I\kern-.18em I}}
\def\IK{\rlx\hbox{\rm I\kern-.18em K}}
\def\IL{\rlx\hbox{\rm I\kern-.18em L}}
\def\one{\hbox{{1}\kern-.25em\hbox{l}}}
\def\0#1{\relax\ifmmode\mathaccent"7017{#1}%
B        \else\accent23#1\relax\fi}

\begin{document}

\begin{titlepage}
\vspace*{-1cm}

\vskip 3cm

\vspace{.2in}
\begin{center}
{\large\bf Self-Dual Skyrmions on the Spheres $S^{2N+1}$ }
\end{center}

\vspace{.5cm}

\begin{center}
Y. Amari~$^{\star,\dagger}$ and L. A. Ferreira~$^{\dagger}$

\vspace{.3 in}
\small

\par \vskip .2in \noindent
$^{\star}$ Department of Physics\\
 Tokyo University of Science\\
  Noda, Chiba 278-8510, Japan

\par \vskip .2in \noindent
$^{\dagger}$ Instituto de F\'\i sica de S\~ao Carlos; IFSC/USP;\\
Universidade de S\~ao Paulo, USP  \\ 
Caixa Postal 369, CEP 13560-970, S\~ao Carlos-SP, Brazil\\

\normalsize
\end{center}

\vspace{.5in}

\begin{abstract}

We construct self-dual sectors for scalar field theories on a $(2N+2)$-dimensional Minkowski space-time with target space being the $2N+1$-dimensional sphere $S^{2N+1}$. The construction of such self-dual sectors is made possible by the introduction of an extra functional on the action that renders the static energy and the self-duality equations conformally invariant on the $(2N+1)$-dimensional spatial submanifold. The conformal and target space symmetries are used to build an ansatz that leads to an infinite number of exact self-dual solutions with arbitrary values of the topological charge. The five dimensional case  is discussed in detail where it is shown that two types of theories admit self dual sectors. Our work generalizes the known results in the three-dimensional case that leads to an infinite set of self-dual Skyrmion solutions.

\end{abstract} 
\end{titlepage}

\section{Introduction}
\label{sec:intro}
\setcounter{equation}{0}

The beauty of self-duality is that it is characterized by first order differential equations such that their solutions also solve the second order Euler-Lagrange equations of the full theory. The self-dual solutions, which in general can be constructed analytically, saturate a lower bound of the energy or Euclidean action, for each sector characterized by the value of the topological charge. The examples include the instantons in Yang-Mills theories in four dimensional Euclidean space \cite{instanton}, the BPS monopoles in three dimensions \cite{bogo,prasad}, the Belavin-Polyakov self-dual solutions of the $O(3)$ or $CP^1$ non-linear sigma model in $(2+1)$-dimensions \cite{bp}, the one-soliton solutions of integrable field theories in $(1+1)$ dimensions like the sine-Gordon model \cite{sg},  field theories for $d$ scalar fields in $(d+1)$-dimensions \cite{adam}, which for the case of $d=3$ include modifications of the Skyrme model \cite{wojtek,shnir,laf,bpsadam}, and so on. 

The interesting fact about the structures of self-duality  that allow the construction of solutions by performing  one integration less,  is not the use of dynamical conservation laws, but the existence in the theory of a topological charge that admits an integral representation. As explained in section 2 of \cite{adam}, one looks for a splitting of the density of topological charge $Q$ as the product of two quantities, let us say 
\be
Q=\int {\cal A}_{\alpha}\,{\widetilde{\cal A}}_{\alpha}
\lab{splittingtopcharge}
\ee
 where $\alpha$ may stand for a set of indices. Being a topological  quantity means that it is invariant under any smooth variations of the fields, and so the relation $\delta Q=0$, provides an identity for the fields which is bilinear in the quantities  ${\cal A}_{\alpha}$ and ${\widetilde{\cal A}}_{\alpha}$. One then introduces the  self-duality equations as  
 \be
 {\cal A}_{\alpha}=\pm \,{\widetilde{\cal A}}_{\alpha}
 \lab{selfdualintro}
 \ee
  It turns out that the bilinear identity coming from the topological charge together with the self-duality equations imply the Euler-Lagrange equation associated with the functional \cite{adam}
  \be
  E=\frac{1}{2}\int\({\cal A}_{\alpha}^2+{\widetilde{\cal A}}_{\alpha}^2\)
  \lab{energyintro}
  \ee
   which can be the static energy or the Euclidean action of the theory. If the functional $E$ is positive definite it then follows automatically a bound given by the topological charge, i.e.  
   \be
   E=\frac{1}{2}\int\({\cal A}_{\alpha}\pm{\widetilde{\cal A}}_{\alpha}\)^2 + \mid Q\mid\geq \mid Q\mid
   \ee
   Note that for a given splitting of the density of topological charge there is the freedom of transforming the quantities ${\cal A}_{\alpha}$ and ${\widetilde{\cal A}}_{\alpha}$ as 
   \be
   {\cal A}_{\alpha}\rightarrow  {\cal A}_{\beta} \,f_{\beta\alpha}\;; \qquad\qquad \text{and}\qquad\qquad
    {\widetilde{\cal A}}_{\alpha}\rightarrow f^{-1}_{\alpha\beta}\, {\widetilde{\cal A}}_{\beta}
    \lab{freedomf}
    \ee
    where $f_{\beta\alpha}$ is an arbitrary invertible matrix. The possibility of introducing such a matrix is what allows the construction of non-trivial self-dual sectors for Skyrme type models \cite{shnir,laf}. In fact, in order to preserve the Lorentz symmetry that quantity is a matrix in the internal indices only, contained in the set of indices $\alpha$. In the cases  considered in this paper $\alpha$ contains only spatial indices and so $f$ will be a scalar function. 
    
    Under the shift \rf{freedomf} the self-duality equations \rf{selfdualintro} become ${\cal A}_{\beta}\,h_{\beta\alpha}=\pm \,{\widetilde{\cal A}}_{\alpha}$ with $h$ being the symmetric invertible matrix $h\equiv f\,f^T$. The topological charge \rf{splittingtopcharge}  remains unchanged but the energy functional \rf{energyintro} becomes $E=\frac{1}{2}\int\({\cal A}_{\alpha}\,h_{\alpha\beta}\,{\cal A}_{\beta}+{\widetilde{\cal A}}_{\alpha}\,h^{-1}_{\alpha\beta}\,{\widetilde{\cal A}}_{\beta}\)$. If one considers the entries of the matrix $h$ as new extra fields, independent of those originally contained in ${\cal A}_{\alpha}$ and ${\widetilde{\cal A}}_{\alpha}$, one observes a very interesting fact. If one varies $E$ w.r.t. to the fields $h$ one gets that $\delta E=0$, for any variation $\delta h$,  if ${\cal A}\,\delta h\,{\cal A}={\widetilde{\cal A}}\,h^{-1}\,\delta h\,h^{-1}\,{\widetilde{\cal A}}$. But that is guaranteed by the new self-duality equations. Therefore, the solutions of the self-duality equations are not only solutions of the Euler-Lagrange equations associated to the fields contained in ${\cal A}_{\alpha}$ and ${\widetilde{\cal A}}_{\alpha}$, but also solve  the Euler-Lagrange equations associated to the fields $h$. 
   
    In the case of Euclidean Yang-Mills theory for instance, one has that ${\cal A}_{\alpha}$ corresponds to the field tensor $F_{\mu\nu}$, and ${\widetilde{\cal A}}_{\alpha}$ to its Hodge dual ${\widetilde F}_{\mu\nu}=\frac{1}{2}\ve_{\mu\nu\rho\sigma}\,F^{\rho\sigma}$. Then the topological charge is the Pontryagin number $Q=\int d^4x\,{\rm Tr}\(F_{\mu\nu}\,{\widetilde F}^{\mu\nu}\)$,  $E$ is the Euclidean action, i.e. $E=\frac{1}{4}\int d^4x\,{\rm Tr}\(F_{\mu\nu}^2\)=\frac{1}{8}\int d^4x\, {\rm Tr}\( F_{\mu\nu}^2+{\widetilde F}_{\mu\nu}^2\)$, and $F_{\mu\nu}=\pm {\widetilde F}_{\mu\nu}$, the well know self-duality equations. 

In this paper we apply the ideas of \cite{adam}, summarized above,  to construct self-dual sectors for field theories in a  $\(2N+2\)$-dimensional Minkowski space-time, with the target space being the $\(2\,N+1\)$-dimensional sphere $S^{2N+1}$. Our results will generalize therefore, in a quite simple way, the results of \cite{shnir,wojtek} for self-dual Skyrmions on $S^3$.  The solitons are static, and since there are no gauge symmetries, the finite energy condition imposes that the fields should go to  fixed constant values  at spatial infinity. Therefore, as long as topology is concerned, one can compactify the space $\IR^{2N+1}$ into the sphere  $S^{2N+1}$, and so  the soliton solutions carry a topological charge given by the winding number of the map $S_{\rm space}^{2N+1}\rightarrow S_{\rm target}^{2N+1}$, which can be evaluated through the  integral
\be
Q_{2N+1}=\frac{2}{(4\pi)^{N+1}}
		\int d^{2N+1}x~
		\varepsilon^{p_1p_2\cdots p_{2N+1}}
		A_{p_1}H_{p_2p_3}H_{p_4p_5}\cdots H_{p_{2N}p_{2N+1}} 
		\lab{topcargegen}
\ee
where we have parameterized the target space with $N+1$ complex fields $Z_a$, $a=1,2,\ldots N+1$, satisfying the constraint $Z^*_a\,Z_a=1$, and have defined the quantities
\be
A_\mu=i\,Z^\dagger \cdot\partial_\mu Z\;;
		\qquad\qquad   Z^\dagger \cdot  Z=1\;;
		\qquad\qquad \mu\,,\nu=0,1,2\ldots 2\,N+1
		\lab{amudef}
\ee
and
\be
H_{\mu\nu}=\pd_\mu A_\nu-\pd_\nu A_\mu
=i\,\(\partial_{\mu}Z^{\dagger}	\cdot \partial_{\nu}Z-\partial_{\nu}Z^{\dagger}	\cdot \partial_{\mu}Z\)
\lab{hmunudef}
\ee
We shall use the metric with signature $(-)$ for the space coordinates and $(+)$ for the time coordinate, i.e.  $ds^2=dx_0^2-dx_i^2$. In addition we take $\ve^{0\,1\,2\ldots 2N+1}=\ve^{1\,2\ldots 2N+1}=1$. Note that even though the target space is $S^{2N+1}$, the target space symmetry group of such theories is not $SO(2N+2)$. The quantities $A_{\mu}$ and $H_{\mu\nu}$, given above, are invariant only under the subgroup $U(N+1)$, where the fields transform as $Z\rightarrow U\,Z$, $U\in U(N+1)$.  

In references \cite{wojtek,shnir} we have considered the case $N=1$, that has led to an infinite number of exact self-dual Skyrmions   on the three dimensional space $\IR^3$, with the fields taking values on the sphere $S^3$, or equivalently on the group $SU(2)$. In this paper we shall consider the case $N=2$, which corresponds to theories in a Minkowski space-time $\IR^{5+1}$, with target space $S^5$. As we show in the section 3 there are basically two ways of splitting the density of topological charges, leading to two different theories. The static sector of those theories are conformally invariant in $\IR^{5}$, and following the method of \cite{babelon}, that leads to an ansatz based on a generalization of the toroidal coordinates for $\IR^{5}$. The ansatz involve three integers associated to the angles of the toroidal coordinates, and the topological charge is the product of those three integers. For both theories we construct an infinite number of exact self-dual soliton solutions. However, for one of the theories those integers are arbitrary and for the other they have to have  equal modulus. 

We then consider the generic case of theories in $(2N+2)$-dimensional Minkowski space-time with target space $S^{2N+1}$. In such cases the number of possibilities of splitting the density of to topological charge is very large, leading to theories which are conformally invariant in $\IR^{2N+1}$. Again that symmetry leads to a toroidal ansatz depending on $N+1$ integers. We consider the case where the splitting leads to a theory that admits and infinite number of self-dual soliton solutions for arbitrary values of those $N+1$ integers. It is worth mentioning that static Skyrmions in seven space dimensions have been obtained from self-dual Yang-Mills in eight Euclidean dimensions \cite{nakamula}  following the Atiyah-Manton construction \cite{atiyah}. Even though the Skyrmion is obtained from a self-dual solution (instanton) it is not a self-dual Skyrmion in seven dimensions.

The paper is organized as follows: in section \ref{sec:s3solutions} we review the results of \cite{shnir} on the cons\-truc\-tion of self-dual Skyrmions on the three dimensional space $\mathbb{R}^3$ with the target space $S^3$. In section \ref{sec:s5solutions} we consider the case of theories in $(5+1)$-dimensions with target space being the five dimensional sphere $S^5$, and  show in detail  how to use the splitting of the topological charge to construct two types of theories admitting self-dual sectors. We then use the conformal and target space symmetries of the self-duality equations to construct infinite sets of exact self-dual solutions for those two types of theories. We then generalize our results  in section \ref{sec:s2n1solutions} to the case of theories in $(2N+2)$-dimensional Minkowski space-time with target space being the $(2N+1)$-dimensional sphere $S^{2N+1}$. Again we construct an infinite set of exact self-dual solutions for one   type of theory coming from a particular choice of the splitting of the topological charge. In section 
\ref{sec:conclusion} we present our conclusions, and in the appendix \ref{app:conformal}  we give the proof of the conformal symmetry of the self-duality equations and in the appendix  \ref{app:topchargeintegral} we solve some integrals relevant for the calculation of the topological charges of the solutions.

\section{Solutions on $S^3$}
\label{sec:s3solutions}
\setcounter{equation}{0}

We begin with a brief review of the work \cite{shnir} on self-dual skyrmions on the three dimensional space $\mathbb{R}^3$ with the target space $S^3$. 
In this case field configurations are characterized by the topological charge $Q\in \pi_3(S^3)=\mathbb{Z}$ given by the integral formula 
\begin{equation}
Q_3=\frac{1}{8\pi^2}\int d^3x~ \varepsilon^{ijk}A_i\,H_{jk}
\lab{3d Chern-Simons}
\end{equation}
where $A_i$ and $H_{ij}$ are defined in \rf{amudef} and \rf{hmunudef} for $N=1$.
We take the splitting of the topological charge density of the form (see \rf{splittingtopcharge})
\begin{equation}
	\mathcal{A}_i\equiv Mf_1A_i\;;
	 \qquad\qquad\qquad {\widetilde{\cal A}}_i\equiv \frac{1}{e\,f_1}\,\ve_{ijk}\, H^{jk}\;;\qquad\qquad i,j,k=1,2,3.
	 \lab{3dsplit}
\end{equation}
where $M$ and $e$ are coupling constants, and $f_1$ is an arbitrary function.
The self-dual equations for such a splitting are 
\begin{equation}
\lambda f_1^2A^i=\varepsilon^{ijk}H_{jk}
\qquad \qquad \text{with} \qquad \quad
\lambda=\pm Me.
\lab{BPS3D}
\end{equation} 
The solutions of the self-duality equations \rf{BPS3D} solve the  Euler-Lagrange equations associated to the following static energy functional  
\be
E=\frac{1}{2}\int d^3x
\left(
M^2 f_1^2A_i^2
+\frac{1}{e^2f_1^2}\left(\varepsilon^{ijk}H_{jk}\right)^2\right)
\lab{energy3D}
\ee
The BPS bound  for such a static energy is given by
\br
E&=&\frac{1}{2}\int d^3x
~\left(
M f_1 A^i
\pm
\frac{1}{e\,f_1}\varepsilon^{ijk}H_{jk}
\right)^2
\mp
\frac{M}{e}\int d^3x~\varepsilon^{ijk}A_iH_{jk}
\geq \frac{8M\pi^2}{e}|Q_3|
\lab{bound 3D}
\er
Using the methods of \cite{babelon} it was constructed in \cite{shnir} an ansatz by exploring the conformal symmetry of the self-duality equations \rf{BPS3D} in the three-dimensional space $\IR^3$ (see appendix \ref{app:conformal}). The ansatz is given by
\begin{equation}
	Z=\(\sqrt{F(z)}e^{i\,n\,\varphi}, \sqrt{1-F(z)}e^{i\,m\,\xi}\)
\end{equation}
where $m$ and $n$ are integers, and $\(z\,,\,\xi\,,\,\vp\)$ are the toroidal coordinates on $\IR^3$ 
\begin{equation}
	x_1=\frac{a}{p}\sqrt{z}\,\cos\varphi, \qquad
	x_2=\frac{a}{p}\sqrt{z}\,\sin\varphi, \qquad
	x_3=\frac{a}{p}\sqrt{1-z}\,\sin\xi
\end{equation}
where
\begin{equation}
	p=1-\sqrt{1-z}\cos\xi\qquad\qquad
	z\in [0,1] \qquad
	\xi,\ \varphi \in [0,2\pi]
\end{equation}
The infinite set of solutions found in \cite{shnir} are given by 
\begin{equation}
	F=\frac{m^2\,z}{m^2\,z+n^2\,(1-z)} \qquad\qquad
	f_1=\sqrt{\frac{2\,p}{|\lambda|\,a}\frac{|m\,n|}{\left[m^2\,z+n^2\,(1-z)\right]}}
	\lab{solution3D}
\end{equation}
where the sign of $\lambda$ is chosen to keep $f_1$ real, i.e. $\mathrm{sign}(\lambda)=-\mathrm{sign}(m\,n)$. The topological charge and static energy for such solutions are given by 
\begin{equation}
	Q_3=-mn\;;\qquad\qquad\qquad E=\frac{8M\pi^2}{e}|mn|
\end{equation}
It turns out \cite{shnir} that the solutions for the cases $m^2=n^2$ present a spherically symmetry energy density, and for the other cases the energy density has only an axial symmetry around the $x_3$-axis. In Fig.1 we show the isosurfaces of the topological charge density (or equivalently energy density) for 
the $Q_3=-4$ cases, i.e. $(m=2,n=2)$, $(m=4,n=1)$ and $(m=1,n=4)$. It is worth noting that for the cases where  $m^2\neq n^2$ cases, the densities have a toroidal inner structure, which at large distances leads to an oblate ($n>m$) or prolate ($n<m$) shape.
Indeed, in the $(m,n)=(4,1)$ case, the outside looks a prolate but the inside is the dumbbell like form. 
In the $(m,n)=(1,4)$ case, the outside looks  oblate but there is a torus shape core.
On the other hand, every isosurface is a sphere in the $(m,n)=(2,2)$ case.
Note that the energy density have the same profile as the topological charge density. 
 
\begin{center}
	\begin{figure*}[t]
		\includegraphics[width=18cm]{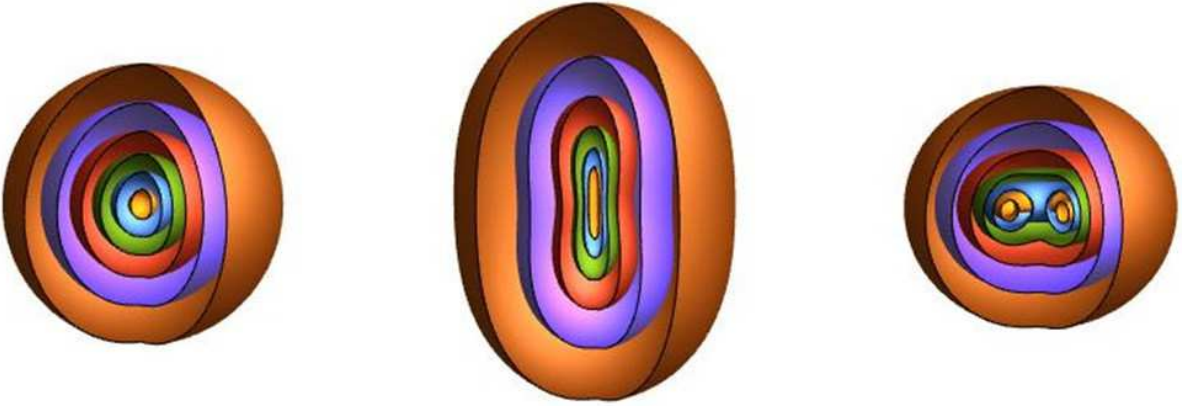}
		\caption{
			The isosurfaces of the topological charge density, from left to right, for $(m,\,n)=(2,\,2)$, $(m,\,n)=(4,\,1)$ and $(m,\,n)=(1,\,4)$.
			 }
		\label{3Dcharge_dens}
	\end{figure*}
\end{center}

\section{Solutions on $S^5$}
\label{sec:s5solutions}
\setcounter{equation}{0}

In this case the topological charge is the winding number of the map $S^5_{\rm space}\rightarrow S^5_{\rm target}$ and given by 
\begin{equation}
   	Q_5=\frac{1}{32\pi^3}\int d^5x~ \varepsilon^{ijklm}A_iH_{jk}H_{lm}
   	\lab{5d Chern-Simons}
   \end{equation}
   There are two basic ways of splitting the density of that topological charge, as in \rf{splittingtopcharge},  to construct theories with exact self-dual sectors as we now explain. 
   
   \subsection{Type I theory on $S^5$} 

 The first case  corresponds to the following splitting of the topological charge density 
\be
{\cal A}^I_i \equiv M\, f_I\, A_i\;; \qquad\qquad\quad {\widetilde{\cal A}}^I_i\equiv \frac{1}{e\,f_I}\,\ve_{ijklm}\, H^{jk}\,H^{lm}\;;\qquad\quad i,j,k,l,m=1,2,\ldots 5
\lab{firstsplit}
\ee
with $A_i$ and $H_{ij}$ defined in \rf{amudef} and \rf{hmunudef} for $N=2$, and  $f_I$ being an arbitrary functional of the complex fields $Z_a$, $a=1,2,3$, and their derivatives, and $M$ and $e$ are coupling constants. Note that the topological charge density does not depend upon the functional $f_I$, and it represents a freedom we have when performing the splitting of it \cite{adam,shnir} (see \rf{freedomf}). The self-duality equation in such a case is 
\begin{equation}
		\lambda f_I^2A^i=\varepsilon^{ijklm}H_{jk}H_{lm}
		\qquad \qquad \text{with} \qquad \quad
		\lambda=\pm Me.
		\lab{BPS1}
\end{equation} 
and solutions of it are solutions of the Euler-Lagrange equations associated to the static energy functional 
\be
E_I=\frac{1}{2}\int d^5x
	 \left(
	 M^2 f_I^2A_i^2
	 +\frac{1}{e^2f_I^2}\left(\varepsilon^{ijklm}H_{jk}H_{lm}\right)^2\right)
	 \lab{energytypeI}
	 \ee
The corresponding action is therefore
\be
S_I=\frac{1}{2}\int d^6x
	 \left(
	 M^2 f_I^2A_{\mu}^2
	 -\frac{1}{2\,e^2f_I^2}\left(\varepsilon^{\mu\nu\rho\sigma\alpha\beta}H_{\rho\sigma}H_{\alpha\beta}\right)^2\right)
	 \lab{actiontypeI}
	 \ee
The bound on the static energy is given by
 \br
		E_I&=&\frac{1}{2}\int d^5x
		~\left(
		M f_I A^i
		\pm
		\frac{1}{e\,f_I}\varepsilon^{ijklm}H_{jk}H_{lm}
		\right)^2
		\mp
		\frac{M}{e}\int d^5x~\varepsilon^{ijklm}A_iH_{jk}H_{lm}
		\notag\\
		&\geq& \frac{32M\pi^3}{e}|Q_5|
		\lab{bound 5DI}
	 \er
	 
	 In order to construct solutions we need an ansatz that explores the external (space) and internal (target)  symmetries of the theory. We shall follow the methods described in \cite{babelon}. As shown in Appendix \ref{app:conformal}, the self-duality equations \rf{BPS1} are invariant under conformal transformations in five dimensions, i.e. it is invariant under the conformal group $SO(6,1)$, which has rank $3$. Therefore, the maximum number of   commuting $U(1)$ subgroups is $3$, and they can be chosen to be generated by the following conformal transformations \cite{babelon}
\br
	&&\pd_{\varphi_1}\equiv x_1\pd_2-x_2\pd_1 \nonumber\\
	&&\pd_{\varphi_2}\equiv x_3\pd_4-x_4\pd_3 
	\lab{spaceu1in5}\\
	&&\pd_\xi\equiv\frac{x_5}{a}
	\left(
	x_1\pd_1+x_2\pd_2+x_3\pd_3+x_4\pd_4
	\right)
	+\frac{1}{2a}\left(
	a^2+x_5^2-x_1^2-x_2^2-x_3^2-x_4^2\right)\pd_5
	\nonumber
	\er
	The first two transformations are infinitesimal rotations on the planes $x_1\textendash x_2$ and  $x_3 \textendash x_4$, and $\varphi_1$ and $\varphi_2$ are the corresponding azimuthal angles. The third transformation is a linear combination of an infinitesimal special conformal transformation $V^{(c_5)}=x_5x_i\pd_i-\frac{1}{2}x_i^2\pd_5$, and an infinitesimal translation $V^{(P_5)}=\pd_5$, along the $x_5$-axis, and $a$ is a free length scale factor. In addition, $\xi$ is the poloidal angle in five dimensions. 
	
The target space symmetries is given by the unitary group $U(3)$, a subgroup of $SO(6)$ which is the symmetry group of $S^5$. Indeed, the operators \rf{amudef} and \rf{hmunudef} are invariant under the transformations 
\be
Z_a\rightarrow U_{ab}\, Z_b\;; \qquad\qquad  Z_a^{*} Z_a=1\;; \qquad\qquad a,b=1,2,3\;;\qquad \qquad U^{\dagger}\cdot U=\one
\ee	
which  has also rank $3$. We shall choose the $3$ (maximum) commuting $U(1)$ subgroups to be 
\be
	\Omega_1=\mathrm{diag}\left(e^{i\alpha_1},1,1\right),
	\qquad
	\Omega_2=\mathrm{diag}\left(1,e^{i\alpha_2},1\right),
	\qquad
	\Omega_3=\mathrm{diag}\left(1,1,e^{i\alpha_3}\right).
	\lab{targetspacerotation}
	\ee	
Following \cite{babelon} we choose an ansatz which is invariant under the joint action of the three external and three internal commuting $U(1)$'s given in 	\rf{spaceu1in5} and \rf{targetspacerotation} respectively. The ansatz is
	\begin{equation}
	Z=
	\left(
	\sqrt{F_1(z, \theta)}~ e^{i n_1 \varphi_1}
	,
	\sqrt{F_2(z, \theta)}~ e^{i n_2 \varphi_2} 
	, 
	\sqrt{1-F_1(z, \theta)-F_2(z, \theta)}~ e^{i m \xi}
	\right)
	\lab{ansatz5D}
	\end{equation}
	where $n_1,n_2$ and $m$ are winding numbers associated the angles $\varphi_1, \varphi_2$ and $\xi$ respectively, and where $z$ and $\theta$ are the two coordinates on $\IR^5$, orthogonal to the three angles $\vp_1$, $\vp_2$, and $\xi$, and defined as 
 \begin{equation}
	z=\frac{4a^2\left(x_1^2+x_2^2+x_3^2+x_4^2\right)}{\left(a^2+x_1^2+x_2^2+x_3^2+x_4^2+x_5^2\right)^2},
	\qquad\qquad
	\theta=\arctan\sqrt{\frac{x_3^2+x_4^2}{x_1^2+x_2^2}}
	\end{equation}
One can check that indeed, $\partial_{\zeta} z=	\partial_{\zeta}\theta=0$, for $\zeta=\(\vp_1\,,\,\vp_2\,,\,\xi\)$. The coordinates $(z,\theta,\xi,\varphi_1,\varphi_2)$, constitute a generalization to $\IR^5$ of the toroidal coordinates on $\IR^3$, and in terms of them the Cartesian coordinates are written   as
	\begin{equation}
	\begin{split}
	& x_1=\frac{a}{p}\sqrt{z}\cos\theta\cos\varphi_1,
	\quad
	x_2=\frac{a}{p}\sqrt{z}\cos\theta\sin\varphi_1,
	\\
	& x_3=\frac{a}{p}\sqrt{z}\sin\theta\cos\varphi_2,
	\quad
	x_4=\frac{a}{p}\sqrt{z}\sin\theta\sin\varphi_2,
	\quad
	x_5=\frac{a}{p}\sqrt{1-z}\sin\xi
	\end{split}
	\end{equation}
	with 
	\begin{equation}
	p=1-\sqrt{1-z}\cos\xi
	\end{equation}
	where the domain of the variables are
	$z\in[0,1],~\theta\in[0,\pi/2],~\xi,\varphi_1,\varphi_2\in[0,2\pi]$.
	In terms of the new coordinates, the metric is written as  
	\begin{equation}
	ds^2=\frac{a^2}{p^2}
	\left(
	\frac{1}{4z(1-z)}dz^2+zd\theta^2+(1-z)d\xi^2+z\cos^2\theta d\varphi_1^2+z\sin^2\theta d\varphi_2^2		
	\right)
	\end{equation}
	
From \rf{amudef}, \rf{hmunudef} and the ansatz \rf{ansatz5D} one observes that $A_{z}=A_{\theta}=0$, and also that $H_{z\theta}=H_{\vp_1\vp_2}=H_{\vp_1\xi}=H_{\vp_2\xi}	=0$. Therefore, the five equations in \rf{BPS1} reduce to only three, since two of them are automatically satisfied by the ansatz \rf{ansatz5D}. In addition, the r.h.s. of \rf{BPS1} for the three remaining equations, are all proportional to the same function of $z$ and $\theta$, namely $\pd_zF_1\pd_\theta F_2-\pd_z F_1\pd_\theta F_2$. Therefore,  substituting the ansatz \rf{ansatz5D} into the BPS equation \rf{BPS1}, leads to the following three  coupled first order partial differential equations 
	\begin{equation}
	\begin{split}
	&\lambda f_I^2\frac{a^3}{p^3}n_1F_1\tan\theta
	=16mn_2
	\left(\pd_zF_1\pd_\theta F_2-\pd_z F_2\pd_\theta F_1\right)
	\\
	&\lambda f_I^2\frac{a^3}{p^3}n_2F_2\cot\theta
	=16mn_1
	\left(\pd_zF_1\pd_\theta F_2-\pd_z F_2\pd_\theta F_1\right)
	\\
	&\lambda f_I^2\frac{a^3}{p^3}\frac{z}{1-z}m(1-F_1-F_2)\sin\theta\cos\theta
	=16n_1n_2\left(\pd_zF_1\pd_\theta F_2-\pd_z F_2\pd_\theta F_1\right)
	\end{split}
	\lab{BPS1explicit}
	\end{equation}
	Since the r.h.s. of \rf{BPS1explicit} are all proportional, they imply that 
		\begin{equation}
		n_1^2F_1\tan\theta
		=n_2^2F_2\cot\theta
		=m^2\frac{z}{1-z}(1-F_1-F_2)\, \sin \theta\,\cos\theta.
		\lab{algebraicequationI}
	\end{equation}
	One can algebraically solve \rf{algebraicequationI} for any non-zero integers $m, n_1$ and $n_2$, and the solutions are given by
	\begin{align}
	\begin{split}
	&F_1=\frac{m^2\,n_2^2\,z\,\cos^2\theta}{n_1^2\,n_2^2\,(1-z)+m^2\,z\,\(n_1^2\,\sin^2\theta+n_2^2\,\cos^2\theta\)},
	\\
	&F_2=\frac{m^2\,n_1^2\,z\,\sin^2\theta}{n_1^2\,n_2^2\,(1-z)+m^2\,z\(n_1^2\,\sin^2\theta+n_2^2\,\cos^2\theta\)},
	\lab{sol typeI}
	\end{split}
	\end{align}
	By substituting such solutions for $F_1$ and $F_2$ into \rf{BPS1explicit}, we obtain
	\begin{equation}
		 f_I=\left(\frac{p^3}{\mid\lambda\mid a^3}\right)^{\frac{1}{2}}
		\frac{4\,\sqrt{2}\,\mid m\,n_1\,n_2\mid^{3/2}}{\left[n_1^2\,n_2^2\,(1-z)+m^2\,z\,\(n_1^2\,\sin^2\theta+n_2^2\,\cos^2\theta\)\right]}.
	\end{equation}
	Since $ f_I$ is a real function, 
	the sign of $\lambda$ and of the integers must satisfy 
	\be
	\text{sign} \,\lambda=\text{sign}\,\(m\,n_1\,n_2\)
	\ee	
The density of the topological charge \rf{5d Chern-Simons} is given by 
\br	
\frac{1}{32\pi^3}\, \varepsilon^{ijklm}A_iH_{jk}H_{lm}&=&
\frac{1}{2\,\pi^3}\,\(\frac{p}{a}\)^5\,\frac{m\,n_1\,n_2}{z\,\sin\theta\,\cos\theta}\,
\left[\partial_zF_1\,\partial_{\theta}F_2-\partial_zF_2\,\partial_{\theta}F_1\right]
\lab{densitytopcharge}
\\
&=&\frac{1}{\pi^3}\,\(\frac{p}{a}\)^5\,\frac{m^5\,n_1^5\,n_2^5}{\left[n_1^2\,n_2^2\,(1-z)+m^2\,z\,\(n_1^2\,\sin^2\theta+n_2^2\,\cos^2\theta\)\right]^3}
\nonumber
\er
where we have used the convention $\ve^{1\,2\,3\,4\,5}=1$, and so  
\be
\ve^{z\,\theta\,\vp_1\,\vp_2\,\xi}=\(\frac{p}{a}\)^5\,\frac{2}{z\,\sin\theta\,\cos\theta}
\lab{epsilondeftoroidal}
\ee
 The volume element is
\be
d^5x=\(\frac{a}{p}\)^5\, \frac{1}{2} \, z\,\sin\theta\,\cos\theta\,dz\,d\theta\,d\vp_1\,d\vp_2\,d\xi
\lab{volumeelement}
\ee 
We now use the fact that
\br
\int_0^1dz\,\int_0^{\frac{\pi}{2}}d\theta\, \frac{ z\,\sin\theta\,\cos\theta}{\left[n_1^2\,n_2^2\,(1-z)+m^2\,z\,\(n_1^2\,\sin^2\theta+n_2^2\,\cos^2\theta\)\right]^3}=\frac{1}{4}\,\frac{1}{m^4\,n_1^4\,n_2^4}
\lab{niceintegral}
\er
to get that the topological charges of those solutions are
\be
Q_5=m\,n_1\,n_2
\ee
For the configurations satisfying the self-duality equations \rf{BPS1} the static energy \rf{energytypeI} becomes
\be
E_I=\int d^5x\;{\cal E}\;;\qquad\qquad\qquad{\rm with}\qquad \qquad{\cal E}=M^2 f_I^2A_i^2
	 \lab{energytypeIselfdual}
	 \ee
The energy density is given by
\be
{\cal E}= 32\, \frac{M}{e} \, \(\frac{p}{a}\)^5\,\frac{\mid m\,n_1\,n_2\mid^5}{\left[n_1^2\,n_2^2\,(1-z)+m^2\,z\,\(n_1^2\,\sin^2\theta+n_2^2\,\cos^2\theta\)\right]^3}
\lab{energydensitytype5d}
\ee
Therefore using \rf{volumeelement} and \rf{niceintegral} one gets
\be
E_I=32\,\pi^3\, \frac{M}{e} \,\mid m\,n_1\,n_2\mid
\ee
From \rf{densitytopcharge} and \rf{energydensitytype5d} one observes that the densities of topological charge and static energy are proportional. In order to visualize the shape of such densities let us write the density of topological charge, given in \rf{densitytopcharge}, in terms of Cartesian coordinates as \begin{equation}
	\mathcal{Q}=\frac{32}{a^5\,\pi^3}
	\frac{\(1+\tilr^2\)(m\,n_1\,n_2)^5}
	{[n_1^2\,n_2^2\(1+\tilr^2\)^2
		+4n_1^2\,\tilrhT^2(m^2-n_2^2)
		+4n_2^2\,\tilrhO^2(m^2-n_1^2)]^3}
\end{equation}
with
\begin{equation}
	\tilrhO=\sqrt{\txa^2+\txb^2},\qquad
	\tilrhT=\sqrt{\txc^2+\txd^2},\qquad
	\tilr=\sqrt{\tilrhO^2+\tilrhT^2+\txe^2}
	\lab{tildecoorddef}
\end{equation}
where $\widetilde{x_i}=x_i/a$.	 Note that $\mathcal{Q}$ does not depend upon the angles $\varphi_1$ and $\varphi_2$, and so the energy and topological charge densities are invariant under the group $SO(2)\times SO(2)$ of rotations on the 
 $x_1$-$x_2$ and $x_3$-$x_4$ planes, for any non-zero values of the integers $m$, $n_1$ and $n_2$. In addition, for the cases where $n_1^2=n_2^2$, such densities depend only upon ${\tilde r}^2$ and ${\tilde x_5}^2$, and so they are invariant under the group $SO(4)$ of rotations on the subspace $\IR^4$ perpendicular do the $x_5$-axis. For the cases where $m^2=n_1^2$ (or $m^2=n_2^2$), the densities depend only upon  ${\tilde r}^2$ and ${\tilde \rho_2}^2$ (or ${\tilde r}^2$ and ${\tilde \rho_1}^2$), and so they are invariant under the group $SO(2)\times SO(3)$ of rotations on the on the plane $x_3$-$x_4$ (or $x_1$-$x_2$), and on the subspace $\IR^3$ perpendicular to the plane $x_3$-$x_4$ (or $x_1$-$x_2$).  Finally for the cases where $m^2=n_1^2=n_2^2$, the densities depend only upon ${\tilde r}^2$ and so they are invariant under the group $SO(5)$ of rotations on the whole space $\IR^5$, i.e. the densities are spherically symmetric. 
 
In Fig.2 we show some examples of surfaces of constant topological charge density in terms of the three coordinates $\tilrhO,\,\tilrhT$ and $\txe$. 
Their structure is very similar to the three dimensional case (see Fig.1).
When $m^2=n_1^2=n_2^2$, the isosurfaces are four dimensional spheres, and so $SO(5)$ invariant.
For $n_1^2=n_2^2$ the iso-surfaces are indeed $SO(4)$ invariant, and note that for $m^2>n_1^2=n_2^2$,  the outer iso-surfaces look a five dimensional prolate but the inside has dumbbell like structure.  
On the other hand, for $m^2<n_1^2=n_2^2$, the outer iso-surfaces look  oblate but the inner shells have a five dimensional torus shape.

\begin{center}
	\begin{figure*}[t]
		\includegraphics[width=16cm]{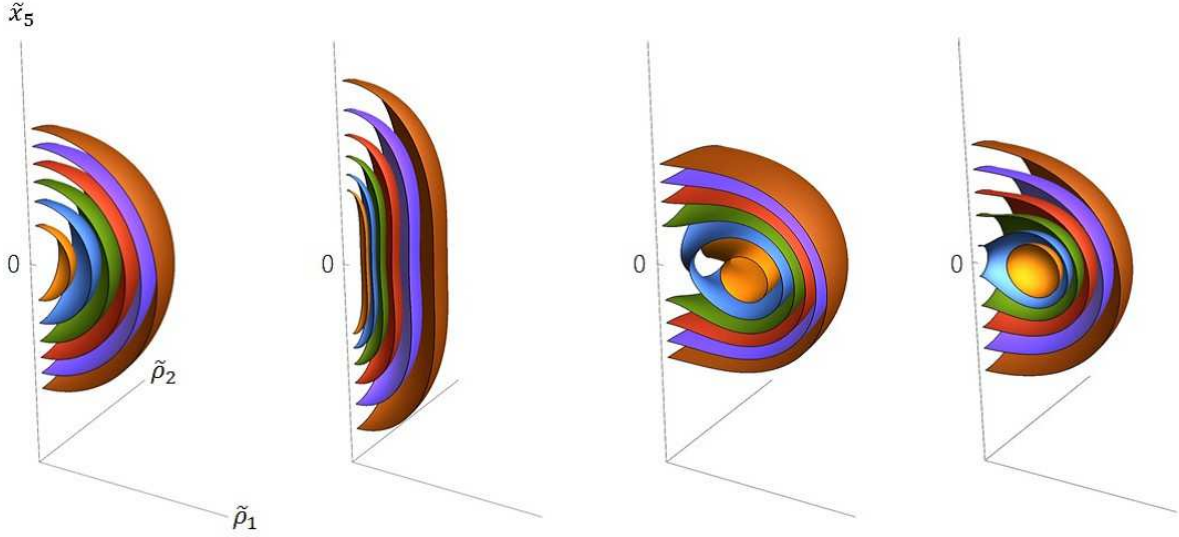}
		\caption{
			The isosurfaces of the topological charge density, from left to right, for 
			$(m,\,n_1,\,n_2)=(1,\,1,\,1)$,
			$(m,\,n_1,\,n_2)=(4,\,1,\,1)$,  
			  $(m,\,n_1,\,n_2)=(1,\,2,\,2)$, and $(m,\,n_1,\,n_2)=(1,\,4,\,1)$ .
		}
		\label{5Dcharge_dens}
	\end{figure*}
\end{center} 
 \subsection{Type II theory on $S^5$} 
 	 
	 The second field theory for the case $N=2$ corresponds to the following splitting of the topological charge density
\be
{\cal A}^{II}_{ij} \equiv M\, f_{II}\, \ve_{ijklm}\,A^k\,H^{lm}\;; \qquad\qquad\qquad {\widetilde{\cal A}}^{II}_{ij}\equiv \frac{1}{e\,f_{II}}\, H_{ij}\;;\qquad\qquad i,j,k,l,m=1,2,\ldots 5
\lab{secondsplit}
\ee	 
with $f_{II}$ having the same nature as $f_I$ introduced above. The self-duality equations in this case are
\be
\lambda\, f_{II}^2\, \ve_{ijklm}\,A^k\,H^{lm}=H_{ij}\qquad \qquad \text{with} \qquad \quad
		\lambda=\pm Me.
		\lab{BPS2}
\ee
Solutions of \rf{BPS2} are also solutions of the Euler-Lagrange equations associated to the static energy functional 
\be
E_{II}=\frac{1}{2}\int d^5x
	 \left(
	 M^2 f_{II}^2\left(\epsilon_{ijklm}\,A^k\,H^{lm}\right)^2
	 +\frac{1}{e^2f_{II}^2}H_{ij}^2\right).
	 \lab{energytypeII}
	 \ee
and the corresponding action is 
\be
S_{II}=-\frac{1}{2}\int d^6x
	 \left(
	 \frac{M^2}{3} f_{II}^2\left(\epsilon_{\mu\nu\rho\sigma\alpha\beta}\,A^{\sigma}\,H^{\alpha\beta}\right)^2
	 +\frac{1}{e^2f_{II}^2}H_{\mu\nu}^2\right).
	 \lab{energytypeII}
	 \ee
The self-duality equation \rf{BPS2} is also invariant under conformal transformations in five dimensions, as shown in the appendix \ref{app:conformal}. Therefore, we shall use the same ansatz, given in \rf{ansatz5D}, used to construct the solutions for the self-duality equations \rf{BPS1}.  When the ansatz \rf{ansatz5D} is replaced into the 10 equations in \rf{BPS2}, one finds that 4 of them are automatically satisfied. The remaining 6 equations are given by 
\br
		\Lambda\, \left(F_1\pd_\theta F_2-F_2\pd_\theta F_1\right)
			&=&-n_1^2\,z(1-z)\,\tan\theta\,\pd_z F_1+\Lambda\, \pd_\theta F_2
			\lab{BPS21a}\\
			\Lambda \,\left(F_1\pd_\theta F_2-F_2\pd_\theta F_1\right)
			&=&-n_2^2\,z(1-z)\,\cot\theta\,\pd_z F_2-\Lambda \,\pd_\theta F_1
			\lab{BPS21b}\\
			\Lambda\, \left(F_1\pd_\theta F_2-F_2\pd_\theta F_1\right)
			&=& m^2\,z^2\,\sin\theta\,\cos\theta\,\pd_z(F_1+F_2)
			\lab{BPS21c}\\
			\Lambda\, \left(F_1\pd_z F_2-F_2\pd_z F_1\right)
			&=& \frac{n_1^2}{4}\,\frac{\tan\theta}{z}\,\pd_\theta F_1+\Lambda \,\pd_z F_2
			\lab{BPS22a}\\
			\Lambda\, \left(F_1\pd_z F_2-F_2\pd_z F_1\right)
			&=&\frac{n_2^2}{4}\,\frac{\cot\theta}{z}\,\pd_\theta F_2-\Lambda\, \pd_z F_1
			\lab{BPS22b}\\			
			\Lambda\, \left(F_1\pd_z F_2-F_2\pd_z F_1\right)
			&=&-\frac{m^2}{4}\,\frac{\sin\theta\cos\theta}{1-z}\,\pd_\theta(F_1+F_2)
			\lab{BPS22c}
	\er
where we have denoted
\be
\Lambda \equiv \lambda\, f^2_{II}\,\frac{p}{a}\,n_1\, n_2\,m
\lab{lambdaconvention}
\ee
The structure of the equations \rf{BPS21a}-\rf{BPS22c} is more complex than that of equations \rf{BPS1explicit}, and we have to analyze them more carefully. Subtracting \rf{BPS21b} from \rf{BPS21a} and then combining with \rf{BPS22c}, multiplied by $\Lambda$, one gets
\be
\left[\Lambda^2\,F_2-\frac{m^2}{4}\,z\,n_1^2\,\sin^2\theta\,\right]\,\partial_zF_1=
\left[\Lambda^2\,F_1-\frac{m^2}{4}\,z\,n_2^2\,\cos^2\theta\,\right]\,\partial_zF_2
\lab{nicebps21}
\ee
Now, subtracting \rf{BPS22b} from \rf{BPS22a} and then combining with \rf{BPS21c}, 
multiplied by $\Lambda$, one gets
\be
\left[\Lambda^2\,F_2-\frac{m^2}{4}\,z\,n_1^2\,\sin^2\theta\,\right]\,\partial_{\theta}F_1=
\left[\Lambda^2\,F_1-\frac{m^2}{4}\,z\,n_2^2\,\cos^2\theta\,\right]\,\partial_{\theta}F_2
\lab{nicebps22}
\ee
Relations \rf{nicebps21} and \rf{nicebps22}  imply that
\be
\left[\Lambda^2\,F_1-\frac{m^2}{4}\,z\,n_2^2\,\cos^2\theta\,\right]\,
\left[\Lambda^2\,F_2-\frac{m^2}{4}\,z\,n_1^2\,\sin^2\theta\,\right]\,
\left[\partial_zF_1\, \partial_{\theta}F_2-\partial_zF_2\, \partial_{\theta}F_1\right]=0
\ee
If we impose $\partial_zF_1\, \partial_{\theta}F_2-\partial_zF_2\, \partial_{\theta}F_1=0$, then it follows that the density of topological charge vanishes (see \rf{densitytopcharge}), and so the solution will be topologically trivial. Therefore, we have to take 
\be
F_1=\frac{m^2\,z}{4\,\Lambda^2}\, n_2^2\,\cos^2\theta\;;\qquad\qquad\qquad\qquad
F_2=\frac{m^2\,z}{4\,\Lambda^2}\, n_1^2\,\sin^2\theta
\lab{funnyf1f2}
\ee
But \rf{funnyf1f2} implies that both $F_1$ and $F_2$ have the same $z$-dependence, and so it follows that $F_1\,\partial_zF_2- F_2\,\partial_zF_1=0$. But from \rf{BPS22c} that implies that $\partial_{\theta}\(F_1+F_2\)=0$, and consequently one has, using \rf{funnyf1f2},  that 
\be
\Lambda^2= \frac{m^2\,z}{4\,\eta\(z\)^2}\,\left[n_1^2\,\sin^2\theta+ n_2^2\,\cos^2\theta\right]
\lab{prelambda}
\ee
for some function $\eta\(z\)$. Therefore,  
\be
F_1=\eta\(z\)^2\, \frac{n_2^2\,\cos^2\theta}{n_1^2\,\sin^2\theta+ n_2^2\,\cos^2\theta}\;;\qquad\qquad
F_2=\eta\(z\)^2\, \frac{n_1^2\,\sin^2\theta}{n_1^2\,\sin^2\theta+ n_2^2\,\cos^2\theta}
\ee
Subtracting \rf{BPS21b} from \rf{BPS21a} and using the relations above, one gets an equation that can only be satisfied if $n_1^2=n_2^2\equiv n^2$. Now, multiply \rf{BPS21a} by $\cos^2\theta$, add to \rf{BPS21b} multiplied by $\sin^2\theta$, and subtract from \rf{BPS21c} to get
\be
\partial_z\eta^2\left[m^2\,z^2+n^2\,z\,\(1-z\)\right]-2\,\eta^2\,\Lambda=0
\lab{finalnice1}
\ee 
Subtracting \rf{BPS22a} from \rf{BPS22b} one gets
\be
2\,z\,\Lambda \partial_z\eta^2-n^2\,\eta^2=0
\lab{finalnice2}
\ee
Multiplying \rf{finalnice2} by $2\,\Lambda$ and subtracting \rf{finalnice1}, multiplied by $n^2$, one gets
\be
 \partial_z\eta^2\left[4\,\Lambda^2-n^2\,\left(m^2\,z+n^2\(1-z\)\right)
 \right]=0
\ee
If we take $\eta$ to be constant, then $F_1$ and $F_2$ do not depend upon $z$, and so
$\partial_zF_1\, \partial_{\theta}F_2-\partial_zF_2\, \partial_{\theta}F_1=0$, which means that the density of topological charge vanishes (see \rf{densitytopcharge}), and we do not want that because the solutions would be topologically trivial. We then have to take $\Lambda^2=n^2\,\left(m^2\,z+n^2\(1-z\)\right)/4$. But to make that compatible with \rf{prelambda}, we need to take 
$\eta^2=m^2\,z/\left[m^2\,z+n^2\(1-z\)\right]$. But replacing that into \rf{finalnice2}, with  $\Lambda=\pm \mid n \mid\,\sqrt{m^2\,z+n^2\(1-z\)}/2$, one gets that we need $\mid n\mid=\pm \sqrt{m^2\,z+n^2\(1-z\)}$. The only possible solution is $m^2=n^2$, and to take $\Lambda$ to be positive, which from \rf{lambdaconvention} one gets the restriction
\be
{\rm sign}\(\lambda\)= {\rm sign} \(n\)
\ee
where we have denoted 
\be
n_1^2=n_2^2=m^2\equiv n^2
\lab{n1n2mnrelation}
\ee
Summarizing, the self-dual solutions are
\br
F_1=z\, \cos^2\theta\;;\qquad\qquad 
F_2=z\, \sin^2\theta\;;\qquad\qquad f_{II}= \sqrt{\frac{1}{2\,\mid \lambda\, n\mid}\,\frac{a}{p}}
\lab{soltype2}
\er

Note that the solutions \rf{soltype2} for $F_1$ and $F_2$, are the same as the solutions \rf{sol typeI} for the cases where $m^2=n_1^2=n_2^2$. Consequently, the solutions for the fields $Z_a$, and so for the vector $A_i$ and tensor $H_{ij}$, are the same for the theory of type II \rf{energytypeII} as for the theory of type I \rf{energytypeI}, for the cases  $m^2=n_1^2=n_2^2$. The solutions for the functions $f_I$ and $f_{II}$ however are different, even if   $m^2=n_1^2=n_2^2$. Since the topological charge density does not depend upon the functions $f_I$ and $f_{II}$, it is the same for those two classes of solutions of such two types of theories.  Therefore, the topological charge for the solutions \rf{soltype2} is given by 
\be
Q_{5}=\mathrm{sign}(m\,n_1\,n_2)\ |n|^3
\ee
where the sign of the charge comes from the choice of relative signs between $n$ and the integers $n_1$, $n_2$ and $m$ in the relation \rf{n1n2mnrelation}.   

The energy densities of those solutions are also the same due to their self-dual character. Indeed,  from \rf{BPS1} and \rf{energytypeI} one obtains that for self-dual solutions one has 
$E_I\sim \int d^5x  f_I^2A_i^2$. Similarly, from \rf{BPS2} and \rf{energytypeII} one gets that $ E_{II}\sim \int d^5x \frac{1}{f_{II}^2}H_{ij}^2$, for self-dual solutions. But \rf{BPS1} implies  $ f_I^2A_i^2\sim \varepsilon^{ijklm}A_iH_{jk}H_{lm}$, and \rf{BPS2} implies $\frac{1}{f_{II}^2}H_{ij}^2\sim \varepsilon^{ijklm}A_iH_{jk}H_{lm}$. Consequently, for the solutions \rf{soltype2}, 
the topological charge density and energy density are proportional, and are spherically symmetric, like the solutions \rf{sol typeI} for the cases where $m^2=n_1^2=n_2^2$ (see discussion below \rf{tildecoorddef}). In fact, we have that the energy of the solutions \rf{soltype2} is given by
\be
E_{II} = \frac{32\,\pi^3\,M}{e}|n|^3
\ee

\section{Solutions on $S^{2N+1}$}
\label{sec:s2n1solutions}
\setcounter{equation}{0}

For the case of self-dual models defined on $\IR^{2N+1}$, with generic values of $N$, there are many possibilities for the splitting of the density of topological charge \rf{topcargegen}. We shall consider only the case where the splitting is such that
\be
{\cal A}^{N}_{p_1}\equiv M\, f_N\, A_{p_1}\;;\qquad\qquad\qquad 
{\widetilde{\cal A}}^{N}_{p_1}\equiv \frac{1}{e\,f_N}\, \varepsilon_{p_1p_2\cdots p_{2N+1}}
		H^{p_2p_3}H^{p_4p_5}\cdots H^{p_{2N}p_{2N+1}} 
		\lab{splittinggeneralN}
		\ee
and the self-duality equation is 
\be
\lambda\,f_N^2\, A_{p_1}=\varepsilon_{p_1p_2\cdots p_{2N+1}}
		H^{p_2p_3}H^{p_4p_5}\cdots H^{p_{2N}p_{2N+1}}\;;
		\qquad \qquad \text{with} \qquad \quad
		\lambda=\pm Me.
		\lab{BPSN}
\ee
Therefore, according to the reasoning explained in the Introduction (section \ref{sec:intro}),  solutions of \rf{BPSN} are solutions of the Euler-Lagrange equations following from the static energy functional given by
\be
E_N= \frac{1}{2}\int d^{2N+1}x\left[ M^2 \,f_N^2\,A_i^2+\frac{1}{e^2\,f_N^2}\(
\varepsilon_{p_1p_2\cdots p_{2N+1}}
		H^{p_2p_3}H^{p_4p_5}\cdots H^{p_{2N}p_{2N+1}}\)^2\right]
		\ee
The corresponding action in the $(2N+2)$-dimensional Minkowski space-time is 
\be
S_N= \frac{1}{2}\int d^{2N+2}x\left[ M^2 \,f_N^2\,A_{\mu}^2-\frac{1}{2\,e^2\,f_N^2}\(
\varepsilon_{\mu_0\mu_1\mu_2\cdots \mu_{2N+1}}
		H^{\mu_2\mu_3}H^{\mu_4\mu_5}\cdots H^{\mu_{2N}\mu_{2N+1}}\)^2\right]
		\ee
The bound on the static energy is given by
\br
E_N&=& \frac{1}{2}\int d^{2N+1}x\left[ M \,f_N\,A_{p_1}\pm \frac{1}{e\,f_N}
\varepsilon_{p_1p_2\cdots p_{2N+1}}
		H^{p_2p_3}H^{p_4p_5}\cdots H^{p_{2N}p_{2N+1}}\right]^2
		\nonumber\\
		&\mp& \frac{M}{e}\,\int d^{2N+1}x\,\varepsilon_{p_1p_2\cdots p_{2N+1}}\,A^{p_1}\,H^{p_2p_3}H^{p_4p_5}\cdots H^{p_{2N}p_{2N+1}}
		\nonumber\\
		&\geq& \frac{\(4\pi\)^{N+1}\,M}{2\, e}\mid Q_{2N+1}\mid
		\er
with $Q_{2N+1}$ given in \rf{topcargegen}. Clearly the bound is saturated by  solutions of the self-duality equations \rf{BPSN}.

In order to construct solutions to the self-duality equations \rf{BPSN} we explore their symmetries. As discussed below  \rf{hmunudef}, the quantities $A_i$ and $H_{ij}$ are invariant under the transformations $Z\rightarrow U\,Z$, with $U\in U\(N+1\)$, and so  \rf{BPSN} are invariant under such $U\(N+1\)$ symmetry. On the other hand, as shown in the appendix \ref{app:conformal}, the self-duality equations \rf{BPSN} are invariant under the conformal group $SO\(2N+2,1\)$. It turns out that both $U\(N+1\)$ and $SO\(2N+2,1\)$ have $N+1$ commuting $U(1)$ subgroups. For the case of $U\(N+1\)$ those subgroups can be taken to form the Cartan subgroup of diagonal matrices, i.e. $U={\rm diag}\(e^{i\alpha_1}\,,\,e^{i\alpha_2}\,,\,\ldots e^{i\alpha_{N+1}}\)$. For the conformal group $SO\(2N+2,1\)$ we shall take those commuting $U(1)$ subgroups to be generated by $N$ commuting spatial rotations plus a linear combination of a special conformal transformation and a translation along the $x_{N+1}$-axis, as follows (see \cite{babelon} for details) 
\br
\partial_{\vp_i}&\equiv& x_{2i-1}\partial_{x_{2i}}-x_{2i}\partial_{x_{2i-1}}\;;
\qquad\qquad\qquad\qquad i=1,2,\ldots N
\nonumber\\
\partial_{\xi}&\equiv&\frac{x_{2N+1}}{a}\,\sum_{i\neq 2N+1}x_i\partial_{x_i}+\frac{1}{2a}\(a^2+x_{2N+1}^2-\sum_{i\neq 2N+1}x_i^2\)\partial_{x_{2N+1}}
\er
with $a$ an arbitrary parameter with dimension of length. We shall construct an ansatz that is invariant under the diagonal action of the internal and external $N+1$ commuting $U(1)$ subgroups, i.e. $e^{i\alpha_i}\otimes \partial_{\vp_i}$, $i=1,2,\ldots N$, and $e^{i\alpha_{2N+1}}\otimes \partial_{\xi}$. The appropriate coordinates for such an ansatz is a generalization of the toroidal coordinates to $\IR^{2N+1}$, made of the angles $\vp_i$, $i=1,2,\ldots N$, and $\xi$, together with  coordinates $z$, $0\leq z\leq 1$, and  $y_{\alpha}$, $\alpha=1,2,\ldots N-1$, with $0\leq y_{\alpha}\leq1$, where the Cartesian coordinates are written as follows   
\br
x_1&=& \frac{a}{p}\,\sqrt{z}\,\sqrt{1-y_1}\,\cos\vp_1\;;\qquad\qquad \qquad \quad
x_2= \frac{a}{p}\,\sqrt{z}\,\sqrt{1-y_1}\,\sin\vp_1	
\nonumber\\
x_3&=& \frac{a}{p}\,\sqrt{z}\,\sqrt{y_1\(1-y_2\)}\,\cos\vp_2\;;\qquad \qquad\quad
x_4= \frac{a}{p}\,\sqrt{z}\,\sqrt{y_1\(1-y_2\)}\,\sin\vp_2
\nonumber\\
x_5&=& \frac{a}{p}\,\sqrt{z}\,\sqrt{y_1\,y_2\,\(1-y_3\)}\,\cos\vp_3\;;\qquad\qquad 
x_6= \frac{a}{p}\,\sqrt{z}\,\sqrt{y_1\,y_2\,\(1-y_3\)}\,\sin\vp_3
\nonumber\\
&\vdots&
\nonumber\\
x_{2\alpha-1}&=&\frac{a}{p}\,\sqrt{z}\,\sqrt{1-y_{\alpha}}\,\prod_{\beta=1}^{\alpha-1}\sqrt{y_{\beta}}\;\,\cos\vp_{\alpha}\;;\qquad \;
x_{2\alpha}=\frac{a}{p}\,\sqrt{z}\,\sqrt{1-y_{\alpha}}\,\prod_{\beta=1}^{\alpha-1}\sqrt{y_{\beta}}\;
\sin\vp_{\alpha}
\nonumber\\
&\vdots&
\nonumber\\
x_{2N-1}&=&\frac{a}{p}\,\sqrt{z}\,\prod_{\alpha=1}^{N-1}\sqrt{y_{\alpha}}\;\cos\vp_N\;;\qquad\qquad \qquad\;
x_{2N}=\frac{a}{p}\,\sqrt{z}\,\prod_{\alpha=1}^{N-1}\sqrt{y_{\alpha}}\;\sin\vp_N
\nonumber\\
x_{2N+1}&=&\frac{a}{p}\,\sqrt{1-z}\, \sin\xi
\er
with $z\in[0,1], y_{\alpha}\in[0,1]$ and $\xi,\vp_i\in[0,2\pi]$, with $\alpha=1,2,\ldots N-1$,  $i=1,2,\ldots N$, and where we have introduced 
\be
p=1-\sqrt{1-z}\cos\xi
\ee
The metric in $\IR^{2N+1}$  is given by 
\be
ds^2=h_{z}^2\, dz^2+\sum_{\alpha=1}^{N-1}h_{y_{\alpha}}^2\,dy_{\alpha}^2+\sum_{i=1}^N h_{\varphi_i}^2\,d\varphi_i^2+h_{\xi}^2\,d\xi^2
\ee
 where the scaling factors are 
	\begin{equation}
	\begin{split}
		&h_{z}=\frac{a}{p}\,\frac{1}{2\sqrt{z\,(1-z)}}\;;
		\qquad
		h_{y_{\alpha}}=\frac{a}{p}\,\sqrt{z}\,\frac{\prod_{\beta=1}^{\alpha-1}\sqrt{y_{\beta}}}{2\sqrt{y_{\alpha}\(1-y_{\alpha}\)}}\;;
		\qquad
		h_{\xi}=\frac{a}{p}\,\sqrt{1-z}\;;
		\lab{scalingfactors}\\
		&h_{\varphi_{\alpha}}=\frac{a}{p}\,\sqrt{z}\,\sqrt{1-y_{\alpha}}\,\prod_{\beta=1}^{\alpha-1}\sqrt{y_{\beta}}\;;
		\qquad
		h_{\varphi_N}=\frac{a}{p}\,\sqrt{z}\,\prod_{\alpha=1}^{N-1}\sqrt{y_{\alpha}}
	\end{split}
	\end{equation}	
with $\alpha=1,2,\ldots N-1$.

The ansatz which is invariant under the diagonal action of the internal and external commuting $U(1)$'s subgroups,  described above, is given by  
	\begin{equation}
		Z=\left(
		\sqrt{F_1(z, y_{\alpha})}~e^{i n_1\varphi_1}
		,
		\sqrt{F_2(z, y_{\alpha})}~e^{i n_2\varphi_2}
		, 
		\cdots
		,
		\sqrt{F_N(z, y_{\alpha})} ~e^{i n_N\varphi_N}
		,
		\sqrt{1-\sum_{k=1}^N F_k}\; \; e^{i m\xi} 
		\right)
		\lab{ansatz2n+1}
	\end{equation}
	where $n_i$ and $m$ are integers. Replacing the ansatz \rf{ansatz2n+1} into the quantities $A_i$ and $H_{ij}$, introduced in \rf{amudef} and \rf{hmunudef}, one obtains that
\begin{equation}
		A_{z}=A_{y_{\alpha}}=0,
		\qquad
		A_\xi=-m\left(1-\sum_{k=1}^N F_k\right),
		\qquad
		A_{\varphi_i}=-n_iF_i,
		\lab{Arelation}
	\end{equation}
and 
\br
H_{\vp_i z}&=&n_i\,\partial_{z}F_i\;;\qquad H_{\vp_i y_{\alpha}}=	n_i\,\partial_{y_{\alpha}}F_i\;;\qquad
H_{\vp_i\vp_j}=H_{\vp_i\xi}=H_{z y_{\alpha}}=H_{y_{\alpha} y_{\beta}}=0
\nonumber\\
H_{\xi z}&=&-m \sum_{k=1}^N \partial_{z}F_k\;;\qquad H_{\xi y_{\alpha}}=-m \sum_{k=1}^N \partial_{y_{\alpha}}F_k
\lab{Hrelation}
	\er
Therefore, from \rf{Arelation} one observes that the l.h.s. of the self-duality equations \rf{BPSN} will be non-zero only when the index $p_1$ corresponds to one of the variables in the set of $N+1$ variables $\(\vp_i\,,\,\xi\)$. On the other hand, the r.h.s. of   \rf{BPSN} contains the product of $N$ components of the tensor $H_{ij}$, and so if $p_1$ does not belong to the set $\(\vp_i\,,\,\xi\)$, the set of indices $p_2\,p_3\ldots p_{2N+1}$ will contain all the indices of that set,   and so at least one of the components of the tensor $H_{ij}$, in that product, will have its two indices in the set $\(\vp_i\,,\,\xi\)$, and so it vanishes. Therefore, both sides  of  \rf{BPSN} vanish when the index $p_1$ does not belong to the set $\(\vp_i\,,\,\xi\)$.  It turns out that when the index $p_1$ belongs to the set $\(\vp_i\,,\,\xi\)$, the  r.h.s. of  \rf{BPSN} will be proportional to $\ve_{p_1r_1r_2\ldots r_N z y_1\ldots y_{N-1}}H^{r_1z}H^{r_2 y_1}\ldots H^{r_N y_{N-1}}$, with the indices $r_i$ taking values in the set $\(\vp_i\,,\,\xi\)$, but different from $p_1$. But that is proportional to the determinant of the $N\times N$ matrix $\partial_{i} F_j$, with the index $i$ belonging to the set of $N$ variables $\(z\,,\, y_{\alpha}\)$. Consequently, the self-duality equations \rf{BPSN} reduce to a set of $N+1$ equations where their l.h.s. is linear in the functions $F_i$, and does not involve their derivatives. On the other hand, their r.h.s. are all proportional to the determinant of the matrix $\partial_{i} F_j$. Choosing the sign of the $\ve$-symbol such that
\be
\ve^{\xi\vp_1\ldots \vp_N z y_1\ldots y_{N-1}}=\frac{1}{h_{\xi}h_{\vp_1}\ldots h_{\vp_N}h_z h_{y_1}\ldots h_{y_{N-1}}}
\ee
one then gets that the self-duality equations \rf{BPSN} imply the  following relations
\br
\frac{n_1^2\,F_1}{h_{\vp_1}^2}&=& \frac{n_2^2\,F_2}{h_{\vp_2}^2}= \ldots = 
\frac{n_N^2\,F_N}{h_{\vp_N}^2}=\frac{m^2}{h_{\xi}^2}\(1-\sum_{k=1}^N F_k\)
\nonumber\\
&=&-\;\frac{\(-1\)^{N(N-1)/2}\,2^N\,N!}{h_{\xi}h_{\vp_1}\ldots h_{\vp_N}h_z h_{y_1}\ldots h_{y_{N-1}}}\, \(m\prod_{k=1}^Nn_k\)\, \frac{{\rm det}\(\partial F\)}{\lambda\,f_N^2}
\lab{niceFrelations}
\er
where
\be
{\rm det}\(\partial F\)\equiv \ve_{i_1i_2\ldots i_N}\,\partial_z F_{i_1}\,
\partial_{y_1} F_{i_2}\,\partial_{y_2} F_{i_3}\,\ldots\,\partial_{y_{N-1}} F_{i_N}
\lab{detdef}
\ee
with $\ve_{1\,2\,3\ldots N}=1$. We are interested in those cases where all the integers $m$ and $n_i$, $i=1,2,\ldots N$, are non-zero since otherwise, as we shown below, the topological charge vanishes. Then, in such cases one can easily solve those algebraic equations to get the $F_i$'s as
\be
F_i=\frac{h_{\vp_i}^2/n_i^2}{h_{\xi}^2/m^2+\sum_{j=1}^Nh_{\vp_j}^2/n_j^2}
\equiv\frac{\kappa_i/n_i^2}{\Delta}\;;\qquad\qquad \Delta=\frac{\kappa_{\xi}}{m^2}+\sum_{j=1}^N\frac{\kappa_j}{n_j^2}\;;\qquad\qquad
i=1,2,\ldots N
\lab{finalfi}
\ee
with
\be
\kappa_{\xi}=\frac{1-z}{z}\;;\qquad \quad
\kappa_N=\prod_{\alpha=1}^{N-1}y_{\alpha}\;;\qquad\quad 
\kappa_{\alpha}=\(1-y_{\alpha}\)\prod_{\beta=1}^{\alpha-1}y_{\beta}\;;\qquad\quad 
\alpha=1,2,\ldots N-1
\lab{kappadef}
\ee
 Therefore, we have that 
$\partial_z F_i= - F_i \frac{\partial_z\Delta}{\Delta}$, and 
$\partial_{y_{\alpha}}F_i=\frac{\partial_{y_{\alpha}}\kappa_i}{n_i^2\,\Delta}-F_i\frac{\partial_{y_{\alpha}}\Delta}{\Delta}$, where $\partial_z\Delta=-1/z^2\,m^2$. Consequently \rf{detdef} becomes
\be
{\rm det}\(\partial F\)=\frac{{\rm det} M}{z^2\,\Delta^{N+1}\,m^2\,\prod_{j=1}^Nn_j^2}
 \ee
where the matrix $M$ has the entries $M_{1j}=\kappa_j$, and $M_{ij}=\partial_{y_{i-1}}\kappa_j$, for $i\geq 2$, and so ${\rm det} M= \ve_{i_1i_2\ldots i_N}\,\kappa_{i_1}\,\partial_{y_1}\kappa_{i_2}\,\partial_{y_2}\kappa_{i_3}\,\ldots \partial_{y_{N-1}}\kappa_{i_N}$. We now introduce the quantities
\be
\iota_N\equiv\sum_{j=1}^N \kappa_j=1\;;\qquad\qquad \iota_{\alpha}\equiv \sum_{j=1}^{\alpha} \kappa_j=1-\prod_{\beta=1}^{\alpha}y_{\beta}\;;\qquad\quad 
\alpha=1,2,\ldots N-1
\ee
Consider a matrix $\varLambda$ with entries $\varLambda_{ij}=1$, for $i\leq j$, and $\varLambda_{ij}=0$, for $i>j$, and so ${\rm det} \varLambda =1$. Therefore, the matrix $N\equiv M\,\varLambda$, has entries $N_{1j}=\iota_j$, and $N_{ij}=\partial_{y_{i-1}}\iota_j$, for $i\geq 2$, and so ${\rm det} M= {\rm det} N=\ve_{i_1i_2\ldots i_N}\,\iota_{i_1}\,\partial_{y_1}\iota_{i_2}\,\partial_{y_2}\iota_{i_3}\,\ldots \partial_{y_{N-1}}\iota_{i_N}$. Since $\iota_N=1$, the only possibility for $i_1$ in that expression is $i_1=N$, and since only $\iota_{N-1}$ depends upon $y_{N-1}$, it follows that the only possibility for $i_N$ is $i_N=N-1$. It then follows that the only possibility for $i_{N-1}$ is $i_{N-1}=N-2$, and so on. Therefore 
\be
{\rm det} M=\ve_{N\, 1\,\,2\,\ldots N-1}\,\partial_{y_1}\iota_1\,\partial_{y_2}\iota_2\,\ldots\,\partial_{y_{N-1}}\iota_{N-1} = y_1^{N-2}\,y_{2}^{N-3}\ldots y_{N-3}^2\, y_{N-2}
\ee
and so
\be
{\rm det}\(\partial F\)=\frac{\prod_{\beta=1}^{N-2}y_{\beta}^{N-\beta-1}}{z^2\,\Delta^{N+1}\,m^2\,\prod_{j=1}^Nn_j^2}
\lab{detMfinal}
 \ee
From \rf{niceFrelations} and \rf{detMfinal} one can determine $f_N$ as
\be
f_N^2= -\,\(\frac{a}{p}\)^2\frac{\(-1\)^{N(N-1)/2}\,2^{N}\,N!}{h_{\xi}h_{\vp_1}\ldots h_{\vp_N}h_z h_{y_1}\ldots h_{y_{N-1}}}\,
\frac{\prod_{\beta=1}^{N-2}y_{\beta}^{N-\beta-1}}{z\,\Delta^{N}\,\lambda\,m\,\prod_{j=1}^Nn_j}
\ee
Since $f_N$ is real one needs 
\be
{\rm sign}\left[\lambda\,m\,\prod_{j=1}^Nn_j\right]=-\(-1\)^{N(N-1)/2}
\lab{signlambdadef}
\ee
Using \rf{scalingfactors} one gets that $h_{\xi}h_{\vp_1}\ldots h_{\vp_N}h_z h_{y_1}\ldots h_{y_{N-1}}=\(\frac{a}{p}\)^{2N+1}\frac{z^{N-1}}{2^N}y_1^{N-2}\,y_{2}^{N-3}\ldots \, y_{N-2}$, and so
\be
f_N=\sqrt{\(\frac{p}{a}\)^{2N-1}\frac{2^{2N}\,N!}{z^N\,\Delta^{N}\,\mid\lambda\,m\,\prod_{j=1}^Nn_j\mid}}
\ee

Let us now evaluate the topological charge $Q_{2N+1}$ given in \rf{topcargegen}. Due to the self-duality equations  \rf{BPSN} one can write it as
\br
Q_{2N+1}=\int d^{2N+1}x \; {\cal Q}_{2N+1}\;;\qquad\qquad
{\cal Q}_{2N+1}=\frac{2\,\lambda}{(4\pi)^{N+1}}\,f_N^2\,A_p^2
\er
Using the solutions given in \rf{Arelation} and \rf{finalfi} one gets that
\be
A_p^2= \frac{A_{\xi}^2}{h_{\xi}^2}+\sum_{i=1}^N\frac{A_{\vp_i}^2}{h_{\vp_i}^2}=\(\frac{p}{a}\)^2\, \frac{1}{z\,\Delta}
\ee
Therefore, the density of topological charge is given by
\be
 {\cal Q}_{2N+1}=\frac{{\rm sign}\(\lambda\)\,2^{2N+1}\,N!}{(4\pi)^{N+1}\,\mid m\,\prod_{j=1}^Nn_j\mid}\,\(\frac{p}{a}\)^{2N+1}\, \frac{1}{z^{N+1}\,\Delta^{N+1}}
 \ee
On the other hand, the volume element is 
\be
d^{2N+1}x=\(\frac{a}{p}\)^{2N+1}\frac{z^{N-1}}{2^N}y_1^{N-2}\,y_{2}^{N-3}\ldots \, y_{N-2}\,
d\xi\,d\vp_1\ldots d\vp_N\,dz\, dy_1\ldots dy_{N-1}
\ee
Integrating in the angles $\xi$ and $\vp_i$, $i=1,2,\ldots N$, one gets that
\be
Q_{2N+1}=\frac{{\rm sign}\(\lambda\)\,N!}{\mid m\,\prod_{j=1}^Nn_j\mid}\int dz\, dy_1\ldots dy_{N-1}\;\frac{y_1^{N-2}\,y_{2}^{N-3}\ldots \, y_{N-2}}{z^{2}\,\Delta^{N+1}}
\lab{topchargeintegral}
\ee
Using the results of the appendix \ref{app:topchargeintegral} (see \rf{finalintegralN}) one gets that
\be
Q_{2N+1}={\rm sign}\(\lambda\)\, \mid m\,\prod_{j=1}^Nn_j\mid=-\(-1\)^{N(N-1)/2}\,m\,\prod_{j=1}^Nn_j
\ee
where we have used the relation \rf{signlambdadef}.

	\section{Conclusions}
	\label{sec:conclusion}
	\setcounter{equation}{0}
	
	 In this paper, we have introduced Skyrme type models  in $(2N+2)$-dimensional Minkowski space-time with the target space being the spheres $S^{2N+1}$. The models do not have a gauge symmetry, and consequently in order to have finite energy static solutions, the fields must go to a constant at spatial infinity. Therefore, as long as topological considerations are concerned, the space sub-manifold $\IR^{2N+1}$ can be  compactified into $S^{2N+1}_{\rm space}$,  and the static solutions define  maps  $S^{2N+1}_{\rm space}\rightarrow S^{2N+1}_{\rm target}$. The topological charge (winding number) associated to such maps has an integral representation, and therefore can be used to construct field theories with self-dual sectors as explained in the introduction.  We have used the freedom described in \rf{freedomf} to introduce an extra functional $f$  that makes the theories conformally invariant in the space sub-manifold $\IR^{2N+1}$. Using the methods of \cite{babelon} we use the conformal group $SO(2N+2,1)$  and the target space symmetry group $U(N+1)$, to construct a static  ansatz based on a generalization of the toroidal coordinates to a space of $\(2N+1\)$ dimensions.  The ansatz was then used to obtain an infinite number of solutions of the self-duality equations carrying non-trivial topological charges. Our construction generalizes the results obtained in \cite{shnir} for the three dimensional case ($N=1$). As shown in  \cite{wojtek,shnir} the three-dimensional models do not present finite energy solutions when the functional $f$ is content. That is a consequence of a theorem due to Chandrasekhar in the context of plasma and solar physics \cite{chandra}. We believe the same happens for the models in $\(2N+1\)$ dimensions considered in this paper, and it would be interesting to generalize that theorem in such more general context. 

As explained in the text the the number of possibilities of splitting the density of the topological charge grows substantially as $N$ increases. Each one of those possibilities leads to a new model. For the five dimensional case ($N=2$) we have considered in detail the two possible models and constructed the topological self-dual skyrmions for them. For the higher dimensional cases ($N>2$) we considered only one possibility corresponding to the case where the self-duality equations imposes the vector $A_i$, defined in \rf{amudef}, multiplied by the functional $f^2_N$, to be proportional to the Hodge dual of the exterior product of $N$ tensors $H_{ij}$, defined in \rf{hmunudef}. That case is physically more interesting because the corresponding theory has a kinetic term quadratic in space-time derivatives of the fields. In addition, it does present restrictions on the possible values of the topological charges of the solutions. 
	
The introduction of the functionals $f_N$, in the splitting of the topological charges, has lead to the conformal symmetry of the models in the space sub-manifold, and made possible the existence of finite energy self-dual solutions of non-trivial topological charges. As we mentioned above, we believe that there can not exist finite energy solutions for such theories with those functional being constants. Despite the important role played by such functionals, their physical nature is not well understood yet, and further studies are necessary to understand them. In addition, it would be interesting to investigate the breaking of the conformal symmetry and its effects on the soliton solutions.

 \vspace{2cm} 
 
 \textbf{Acknowledgements}\\
 
   The authors are very grateful to Nobuyuki Sawado and Kouichi Toda for many helpful discussions.
   YA would like to thank the kind hospitality at the Instituto
   de F\'isica de S\~{a}o Carlos.
   His stay in Brazil is supported by Tokyo University of Science. LAF is partially supported by CNPq-Brazil.

	\appendix
	\section{Conformal symmetry of the BPS equations}
	\label{app:conformal}
	\setcounter{equation}{0}

	In this appendix, we prove that the BPS equations are invariant under the conformal group on $\IR^{2N+1}$. Let us consider infinitesimal coordinate transformations $\delta x_i=\zeta_i$, $i=1,2,\ldots 2N+1$. 	We take the vector of complex fields $Z_a$, $a=1,2,\ldots N+1$, introduced in \rf{amudef},  as scalar fields under such space transformations. Therefore one has \cite{babelon} 
	\begin{equation}
		\delta Z=0,\qquad
		\delta A_i=-\pd_i\zeta_jA_j,\qquad 
		\delta H_{ij}=-\pd_i\zeta_kH_{kj}-\pd_j\zeta_kH_{ik}.
		\lab{conformaltranfdef}
	\end{equation}  
	
The splitting of the topological charge \rf{topcargegen} (see \rf{splittingtopcharge}) leads to self duality equations of the form $\lambda\, f^2\, w= {\tilde v}$, 
where $w$ and $v$ are differential forms constructed out of the vetor and tensor fields $A_i$ and $H_{ij}$, introduced in 	\rf{amudef} and \rf{hmunudef}. One has that $w$ is a $\(2\,p+1\)$-form as  $w=A\wedge H\wedge H\ldots \wedge H$ (with $p$ $H$'s) and $v$ is a $2\(N-p\)$-form as $v=H\wedge H\ldots \wedge H$ (with $\(N-p\)$ $H$'s). In addition,  ${\tilde v}$ is the Hodge dual of $v$. In components, the self-duality equations read 
\be
\lambda\,f^2\, w_{i_1i_2\ldots i_{2p+1}}= \frac{1}{2\(N-p\)!}\,\ve_{i_1i_2\ldots i_{2p+1}j_1j_2\ldots j_{2\(N-p\)}}\, v_{j_1j_2\ldots j_{2\(N-p\)}}
\lab{selfdualappendix}
\ee
with $\lambda=\pm Me$, (see \rf{BPSN}). Using \rf{conformaltranfdef} one then gets that the self-duality equations \rf{selfdualappendix} transform as
\br
&&\lambda\,f^2\left[ 2\frac{\delta f}{f}\, w_{i_1i_2\ldots i_{2p+1}}-\partial_{i_1}\zeta_k\, w_{k i_2\ldots i_{2p+1}}-\partial_{i_2}\zeta_k\, w_{i_1k\ldots i_{2p+1}}\ldots - \partial_{i_{2p+1}}\zeta_k\, w_{i_1i_2\ldots k}\right]=
\nonumber\\
&=&- \frac{\ve_{i_1i_2\ldots i_{2p+1}j_1j_2\ldots j_{2\(N-p\)}}}{2\(N-p\)!}\left[
\partial_{j_1}\zeta_k\,v_{k j_2\ldots j_{2\(N-p\)}}+\partial_{j_2}\zeta_k\,v_{j_1 k\ldots j_{2\(N-p\)}}+\partial_{j_{2\(N-p\)}}\zeta_k\,v_{j_1j_2\ldots k}\right]
\nonumber\\
&=&- \frac{\ve_{i_1i_2\ldots i_{2p+1}j_1j_2\ldots j_{2\(N-p\)}}}{\left[2\(N-p\)-1\right]!}\;
\partial_{j_1}\zeta_k\,v_{k j_2\ldots j_{2\(N-p\)}}
\\
&=&- \frac{\ve_{i_1i_2\ldots i_{2p+1}j_1j_2\ldots j_{2\(N-p\)}}\,\ve_{l_1l_2\ldots l_{2p+1}k j_2\ldots j_{2\(N-p\)}}}{\left[2\(N-p\)-1\right]!\,\(2p+1\)!}\,
\lambda\,f^2\,\partial_{j_1}\zeta_k\,w_{l_1l_2\ldots l_{2p+1}}
\nonumber
\er
where in the last equality we have used the Hodge dual of \rf{selfdualappendix}. Therefore, in order for the self-duality equations to be invariant one needs that
\br
\left[2\frac{\delta f}{f}+\partial_k\zeta_k\right] w_{i_1i_2\ldots i_{2p+1}}-\(\partial_{i_1}\zeta_k+\partial_k\zeta_{i_1}\)\, w_{k i_2\ldots i_{2p+1}}\ldots - \(\partial_{i_{2p+1}}\zeta_k+\partial_k\zeta_{i_{2p+1}}\)\, w_{i_1i_2\ldots k}=0
\nonumber
\er
Such a relations hold true if the space transformations are conformal, i.e. if the functions $\zeta_i$ satisfy
\be
\partial_i\zeta_j+\partial_j\zeta_i=2\,D\,\delta_{ij}
\lab{conformaltransfdef}
\ee
for some function $D$, and if the transformation of the function $f$ satisfies
\be
\delta f= \frac{1}{2}\left[4\,p-2\,N+1\right]\,D\,f
\lab{generalftransform}
\ee
As is shown in \cite{babelon}, \rf{conformaltransfdef} are actually the equations which define the conformal transformations. Indeed, if  $D$ is a linear function of $x_i$, 
	$\zeta_i$ corresponds to the special conformal transformations, 
	if $D$ is a constant, then $\zeta_i$ leads to the dilatations, 
	and if $D=0$, then $\zeta_i$ defines the translations and rotations.
	
		In addition, one can check that 
			\begin{align}
		\delta\left(d^{2N+1}x\right)&=
		(2N+1)\,D\,d^{2N+1}x
		\\
		\delta\left( f^2\, w^2\right)
		&=-(2N+1)\,D \,f^2\,w^2
		\\
		\delta\left( f^{-2}\,{\tilde v}^2\right)
		&=-(2N+1)\,D \,f^{-2}\,{\tilde v}^2
		\\
		\delta\left(w\,{\tilde v}
		\right)
		&=-(2N+1)\,D\, w\,{\tilde v}
	\end{align}
Therefore, the topological charge, $Q\sim \int d^{2N+1}x\; w\,{\tilde v}$, and the static energy, given by  $E\sim \int d^{2N+1}x\left[ M^2\, f^2\,w^2+ \frac{{\tilde v}^2}{e^2\,f^2}\right]$, with $\lambda=\pm M\,e$, are invariant under the conformal transformations in $\IR^{2N+1}$.	

Note that the functions $f_1$, $f_I$ and $f_N$, introduced in \rf{3dsplit}, \rf{firstsplit} and \rf{splittinggeneralN} respectively, correspond the cases $p=0$ and $N=1$, $N=2$ and $N=N$ respectively. The function $f_{II}$, introduced in \rf{secondsplit}, corresponds to the case $p=1$ and $N=2$. Therefore, from \rf{generalftransform}, one has that such functions transform under the conformal group as
\be
\frac{\delta f_1}{f_1}=-\frac{D}{2}\;;\qquad \frac{\delta f_I}{f_I}=-\frac{3}{2}\,D\;;\qquad \frac{\delta f_{II}}{f_{II}}=\frac{D}{2}\;;\qquad \frac{\delta f_N}{f_N}=-\frac{\(2N-1\)}{2}\,D
\ee

	\section{The topological charge integral}
	\label{app:topchargeintegral}
	\setcounter{equation}{0}
		
In this appendix we evaluate  the integral appearing in the expression \rf{topchargeintegral} for the topological charge. In fact, instead of evaluating it directly we find a recursive relation for such integrals. We start with the first one, corresponding to the case $N=1$, and given by
\be
I_1\(m\,,\,n_1\)\equiv \int_0^1\frac{dz}{z^2}\, \frac{1}{\Delta_{(1)}^2}= m^2\,n_1^2\;;\qquad{\rm with}\qquad\qquad 
\Delta_{(1)}\equiv \frac{1-z}{z\,m^2}+\frac{1}{n_1^2}
\lab{integral1}
\ee
The second integral is 
\be
I_2\(m\,,\,n_1\,,\,n_2\)\equiv \int_0^1\frac{dz}{z^2}\,\int_0^1dy_1\, \frac{1}{\Delta_{(2)}^3}\;;\qquad{\rm with}\qquad\qquad 
\Delta_{(2)}\equiv \frac{1-z}{z\,m^2}+\frac{1-y_1}{n_1^2}+\frac{y_1}{n_2^2}
\lab{integral2}
\ee
The quantity $\Delta_{(2)}$ in the denominator is linear in $y_1$ and so the $y_1$-integration can be easily performed to give
\br
I_2\(m\,,\,n_1\,,\,n_2\)=\frac{1}{2}\,\frac{n_1^2\,n_2^2}{\(n_1^2-n_2^2\)}\left[ I_1\(m\,,\,n_1\)-I_1\(m\,,\,n_2\)\right]=\frac{1}{2}\,m^2\,n_1^2\,n_2^2
\er
We now consider the integral appearing in \rf{topchargeintegral} for $N\geq 3$, and given by
\be
I_N\(m\,,\,n_1\,,\,n_2\ldots \,,\,n_{N}\)\equiv \int_0^1\frac{dz}{z^2}\,\int_0^1dy_1\,\ldots \int_0^1dy_{N-1}\,\frac{y_1^{N-2}\,y_2^{N-3}\ldots y_{N-2}}{\Delta_{(N)}^{N+1}}
\lab{integralN}
\ee
where $\Delta_{(N)}$ is the same as $\Delta$ defined in \rf{finalfi}, and that we write here as 
\be
\Delta_{(N)}\equiv \Delta_{(N-1)}+b_{(N)}\,y_{N-1}
\ee
with
\be
\Delta_{(N-1)}\equiv 	\frac{1-z}{z\,m^2}+\sum_{j=1}^{N-2}\frac{\kappa_j}{n_j^2}+
\frac{1}{n_{N-1}^2}\prod_{\beta=1}^{N-2}y_{\beta}\;;\qquad\qquad
b_{(N)}\equiv \left[\frac{1}{n_N^2}-\frac{1}{n_{N-1}^2}\right]\,\prod_{\beta=1}^{N-2}y_{\beta}
\ee
with $\kappa_j$ defined in \rf{kappadef}. Again $\Delta_{(N)}$ is linear in $y_{N-1}$ and the $y_{N-1}$-integration leads to the recursion relation
\br
I_N\(m\,,\,n_1\,,\,n_2\ldots \,,\,n_{N}\)&=&\frac{1}{N}\frac{\(n_{N-1}^2-n_N^2\)}{n_{N-1}^2\,n_N^2}\,
\left[ I_{N-1}\(m\,,\,n_1\,,\,n_2\ldots \,,\,n_{N-2}\,,\,n_{N-1}\)
\right.
\nonumber\\
&-& \left. I_{N-1}\(m\,,\,n_1\,,\,n_2\ldots \,,\,n_{N-2}\,,\,n_{N}\)\right]
\er
Using such a recursion relation one gets that
\be
I_N\(m\,,\,n_1\,,\,n_2\ldots \,,\,n_{N}\)=\frac{1}{N!}\;m^2\,n_1^2\,n_2^2\ldots n_N^2
\lab{finalintegralN}
\ee


\newpage	
		
		
		
	\begin{thebibliography}{99}	
	
	\bibitem{instanton} 
  A.~A.~Belavin, A.~M.~Polyakov, A.~S.~Schwartz and Y.~S.~Tyupkin; 
  ``Pseudoparticle Solutions of the Yang-Mills Equations''; 
  Phys.\ Lett.\  {\bf 59B}, 85 (1975);
  doi:10.1016/0370-2693(75)90163-X\\
  %%CITATION = doi:10.1016/0370-2693(75)90163-X
  M.~F.~Atiyah, N.~J.~Hitchin, V.~G.~Drinfeld and Y.~I.~Manin;
  ``Construction of Instantons''; 
  Phys.\ Lett.\ A {\bf 65}, 185 (1978); 
  doi:10.1016/0375-9601(78)90141-X
  %%CITATION = doi:10.1016/0375-9601(78)90141-X;%%
  
  \bibitem{bogo} E.B. Bogomolnyi, ````The stability of Classical Solutions''
{\it Sov. J. Nucl. Phys.} {\bf 24} 449, 1976.

\bibitem{prasad} M. K. Prasad, C. M. Sommerfield, Phys. Rev. Lett. {\bf 35} (1975) 760.

 \bibitem{bp}
A.A. Belavin and A.M. Polyakov, {\it JETP Lett.} {\bf 22} (1975) 
245-247.
  
  \bibitem{sg} 
	D. Finkelstein and C.W. Misner, {\em Annals of Physics} {\bf 6}, 230 (1959); T.H.R. Skyrme, {\em Proc. Royal Soc.}  {\bf A247} 260 (1958); J.K. Perring and T.H.R. Skyrme, {\em Nucl. Phys.} {\bf 31} 550 (1962); U. Enz, {\em Phys. Rev.} {\bf 131} 1392 (1963); J. Rubinstein, {\em Jour. Math. Phys.} {\bf 11} 258 (1970).
	
	\bibitem{adam} 
  C.~Adam, L.~A.~Ferreira, E.~da Hora, A.~Wereszczynski and W.~J.~Zakrzewski;
  ``Some aspects of self-duality and generalised BPS theories''; 
  JHEP {\bf 1308}, 062 (2013); 
  doi:10.1007/JHEP08(2013)062; 
  [arXiv:1305.7239 [hep-th]].
  %%CITATION = doi:10.1007/JHEP08(2013)062;%%
  
   \bibitem{wojtek} 
  L.~A.~Ferreira and W.~J.~Zakrzewski; 
  ``A Skyrme-like model with an exact BPS bound,''
  JHEP {\bf 1309}, 097 (2013); 
  doi:10.1007/JHEP09(2013)097; 
  [arXiv:1307.5856 [hep-th]].
  %%CITATION = doi:10.1007/JHEP09(2013)097;%%
	
	\bibitem{shnir} 
  L.~A.~Ferreira and Y.~Shnir; 
  ``Exact Self-Dual Skyrmions''; 
  Phys.\ Lett.\ B {\bf 772}, 621 (2017); 
  doi:10.1016/j.physletb.2017.07.040; 
  [arXiv:1704.04807 [hep-th]].
  %%CITATION = doi:10.1016/j.physletb.2017.07.040;%%
  
   \bibitem{laf} 
  L.~A.~Ferreira; 
  ``Exact self-duality in a modified Skyrme model''; 
  JHEP {\bf 1707}, 039 (2017); 
  doi:10.1007/JHEP07(2017)039; 
  [arXiv:1705.01824 [hep-th]].
  %%CITATION = doi:10.1007/JHEP07(2017)039;%%
  
  \bibitem{bpsadam}
C. Adam, J. Sanchez-Guillen, A. Wereszczynski,
Phys. Lett. B{\bf 691}, 105 (2010);
[arXiv:1001.4544];
Phys. Rev. D{\bf 82}, 085015 (2010);   
[arXiv:1007.1567].

\bibitem{babelon} 
  O.~Babelon and L.~A.~Ferreira; 
  ``Integrability and conformal symmetry in higher dimensions: A Model with exact Hopfion solutions'';  
  JHEP {\bf 0211}, 020 (2002); 
  doi:10.1088/1126-6708/2002/11/020; 
  [hep-th/0210154].
  %%CITATION = doi:10.1088/1126-6708/2002/11/020;%%
  
 

\bibitem{nakamula} 
  A.~Nakamula, S.~Sasaki and K.~Takesue,
  ``Atiyah-Manton Construction of Skyrmions in Eight Dimensions,''
  JHEP {\bf 1703}, 076 (2017); 
  doi:10.1007/JHEP03(2017)076; 
  [arXiv:1612.06957 [hep-th]].
  %%CITATION = doi:10.1007/JHEP03(2017)076;%%
  
  \bibitem{atiyah} 
  M.~F.~Atiyah and N.~S.~Manton,
  ``Skyrmions From Instantons,''
  Phys.\ Lett.\ B {\bf 222}, 438 (1989).
  doi:10.1016/0370-2693(89)90340-7
  %%CITATION = doi:10.1016/0370-2693(89)90340-7;%%
  
   \bibitem{chandra}S.~Chandrasekhar, {\it Hydrodynamic and hydromagnetic stability},
Dover Publication, Inc. (1981).

	
	\end{thebibliography}
\end{document}